\documentclass[pre,singlecolumn,showpacs,superscriptaddress] {revtex4}

\usepackage{amsmath,amssymb}
\usepackage{graphics,graphicx}
\usepackage[normalem]{ulem}
\newcommand{\cu}

\begin{document}

\title
{An empirical analysis of the Ebola outbreak in West Africa}

\author{Abdul Khaleque}%
\email[Email: ]{aktphys@gmail.com}
\affiliation{Department of Physics, University of Calcutta, 92 APC Road, Kolkata 700009, India},\affiliation{School of Physical Sciences, National Institute of Science Education and Research
(NISER), Khordha, Jatni, Odisha 752050, India}

\author{Parongama Sen}%
\email[Email: ]{parongama@gmail.com}
\affiliation{Department of Physics, University of Calcutta, 92 APC Road, Kolkata 700009, India}

\begin{abstract}

The data for the Ebola outbreak that occurred in 2014-2016 in three countries
of West Africa are analysed within a common framework. The
analysis is made using the results of an agent based
Susceptible-Infected-Removed (SIR) model on a Euclidean network, where nodes at a distance $l$ are
connected with probability $P(l) \propto l^{-\delta }$, $\delta$ determining the range of the interaction, in addition
to nearest neighbors. The cumulative (total) density of infected 
population here has the form $R(t) = \frac{a\exp(t/T)}{1+c\exp(t/T)}$, where the parameters depend on $\delta$ and 
the infection probability $q$.
This form is seen to fit well with the data. 
Using the best fitting parameters, the time at which the peak is reached is  estimated and is shown to be consistent with the data.
We also show that in the Euclidean model, one can choose $\delta$ and $q$ values which reproduce
the data for the three countries qualitatively. 
These choices are correlated with  population density, control schemes and other factors.
Comparing the real data and the results from the model one can also estimate 
the size of the actual population susceptible to the disease. 
Rescaling the real data a reasonably good quantitative agreement with the simulation results is obtained.

\end{abstract}
\pacs{87.19.Xx,07.05.Kf, 95.75.-z}
\maketitle


\section{Introduction}

Mathematical modelling  of the phenomena of disease spreading   
has a long history, the first such attempts being  made in  the 
early twentieth century \cite{Ross:1916,Ross:1917,Brownlee:1918,Kermack:1927,Soper:1929,Kermack:1932,Kermack:1933}.
 Typically, an individual is assumed to be in   either one of the three possible states:  
 susceptible, infected and   removed (or recovered) denoted by S, I, and R respectively in the simplest models.
Diseases which can be contracted only once are believed to be described by the SIR model in which a susceptible 
individual gets infected by an infected agent
who is subsequently  removed (dead or recovered). A removed person no longer takes part in the dynamics. 
In SIS model, an infected person may become susceptible again. 
   In the SIR model, $S,I$ and $R$ represent the densities of  population in the three different states and  
  are related through the normalization
condition 
\begin{equation}
S(t)+I(t)+R(t)=1,
\label{norm}
\end{equation}
The following set of deterministic differential equations are obeyed by the densities:
\begin{equation}
\frac{dS}{dt} = -q(k-1)IS, \\ 
\label{rates1}
\end{equation}
\begin{equation}
\frac{dI}{dt} =-{\mu}I +q(k-1)IS, \\ 
\label{rates2}
\end{equation}
\begin{equation}
\frac{dR}{dt} ={\mu}I. 
\label{rates3}
\end{equation}
These equations can be interpreted as follows: infected nodes become recovered at a rate $\mu$, while susceptible nodes become infected at a rate 
proportional to both the densities of infected and susceptible nodes. Here, $q$ is the infection rate and $k$ is the number of 
contacts or degree.  
Without loss of generality, one can take $\mu =1$. 
 Due to the conservation of the total population (eq. \ref{norm}), only two of the three variables are independent. 
For the SIS model, one has only two similar equations connecting $S$ and $I$, one of which is independent only. 
 In most
theoretical models, the epidemic has a threshold behaviour as the 
infection probability $q$ is varied. However, an estimate of $q$ from 
real data 
is difficult as it is  related to biological features like nature of the 
pathogen etc.

Plenty of variations and modifications
  of the SIR and SIS models have been considered  
 over the last few decades. Resurgence of interest in these models has taken place following
the discovery that  social networks do not behave like random or regular 
networks \cite{Barrat:2008,Sen:2013}. The  current  emphasis has been  to study these models on 
complex networks like small world and scale free networks.  A few surprising results have been
derived theoretically in the recent past \cite{Sen:2013}.

The test of a model lies in its ability to match real data. No appreciable success 
has been made so far for the familiar models although some qualitative consistency has been 
achieved \cite{Sen:2013,Hethcote:2000}. 
The available data is usually  in the form of number of newly infected patients and total (cumulative) number of cases.  
In  the SIR model, the newly infected fraction shows an initial growth followed by a peak and a subsequent 
 decay. This matches with the overall structure of the real data (e.g. for Severe Acute Respiratory Syndrome (SARS) \cite{Watts:2005}), which however, 
 show local
oscillatory behaviour in addition. Such a behaviour may be due to demographic non uniformity \cite{Hethcote:1974}. 

The set of equations (\ref{rates1},\ref{rates2},\ref{rates3}) represent only a mean field picture. 
The mean field equations do not 
depend on the topology of the network and are also essentially deterministic. 
It is therefore more meaningful 
to study  the epidemic spreading by considering an agent based model on spatial networks where the dynamics of each agent can 
be tagged and the averages can be extracted easily. 
Agent based models for epidemic spreading  on regular lattices have been studied quite extensively in the last few decades 
and in the more recent studies, the   complex nature of the network connecting the individuals has been taken into  consideration. 
It has been shown that the geographical factor plays an important role in the spreading 
process \cite{Janssen:1999,Linder:2008,Munoz:2010,Wu:2004,Xu:2006,Xu1:2007,Zhao:2012,Khaleque:2013,Grass:2013}.
In particular, the SIR model on an Euclidean network, where the agents may be connected not only to their nearest neighbours but also to a 
few randomly chosen long range neighbours
has been considered in 
 detail \cite{Khaleque:2013,Grass:2013}. 

In 2014,  the Ebola virus caused large scale outbreaks mainly in three West African countries and only 
 recently it has been declared as over (June 2016). 
Ebola virus  is transmitted through body fluids and it is also believed that a person
can contract the disease only once. 
A few attempts have been made to analyse the data so far \cite{Lew:2014,Chow:2015,Cam:2014,Amira:2015,Red:2015,Burg:2015}. 
Different factors like demographic effect, 
hospitalization, vaccination and  treatment plans have been 
incorporated in  the  traditional and well-known SIR model to understand the dynamics of Ebola disease \cite{Amira:2015,Red:2015,Burg:2015}.
However, in these models, a mean field approximation has been used which is rather unphysical.
Using the results of an agent based SIR model on Euclidean network 
 \cite{Khaleque:2013}, 
mentioned in the last paragraph,
we have analysed the Ebola data   
 for the three countries Guinea, Liberia and Sierra Leone in West Africa where the outbreak
extended over approximately two years. We have also reproduced the comparative treads using appropriate parameter values
in the model, albeit qualitatively.

\section{Results}


\subsection{Data analysis}
We have studied the available
data for total (cumulative) number of cases  $R(t)$ as a function time $t$ and  extracted the data for  number of new cases $I(t)$  from these.

\begin{figure}[ht]
\centering
\includegraphics[width=9.5cm]{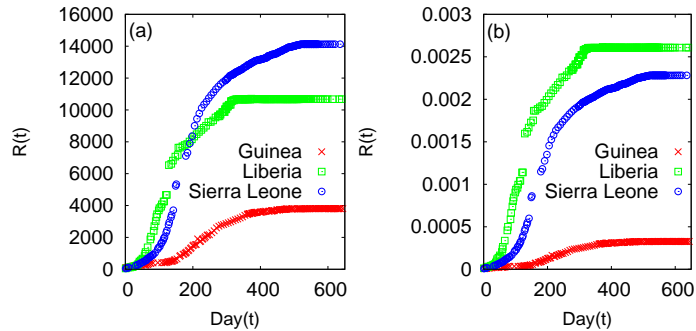}
\caption{(a) Cumulative number of infected individuals as a function of time (day) for the three countries Guinea, Liberia and
Sierra Leone. (b)  Same data normalised by the population of each country.}
\label{fig:gls}
\end{figure}

Most of the earlier studies have  dealt with the actual numbers of cases. However, as we attempt to provide a comparative picture, 
we have taken the fraction i.e., divided the numbers by the total population $N_p$ for each country. One can easily see that the comparative trends become
 different in the two different approaches (Figure \ref{fig:gls}). 
The disease is seen to affect the least fraction 
of the population in Guinea and the maximum in Liberia. However, the number of cases is maximum for Sierra Leone and not for Liberia.
Considering the  time
at which the data reach a saturation value, one can also conclude that the disease has existed over a   longer  period in the case of Sierra Leone and Guinea.

Had the infection probability been the sole factor responsible for the spread, 
the patterns would have been the same for the three countries.
We argue that the network structure  is responsible for the different trends.
Hence   a theoretical model that yields  results comparable to the observed data   
must have more than one parameter. 
A minimal model would consist of two parameters like the one considered in \cite{Khaleque:2013}. Here the agents have two  nearest neighbour 
connections and a random long range connection to a agent located at a distance $l$ with probability $ l^{-\delta}$ (details given in the Methods Section).
 The parameter $\delta$ essentially  controls the network structure and 
the other parameter is of course the infection probability $q$. This study revealed that for a given $\delta$, above a 
threshold value of $q$ (which depends on $\delta$), an epidemic can occur.

\begin{figure}[ht]
\centering
\includegraphics[width=13.5cm,angle=0]{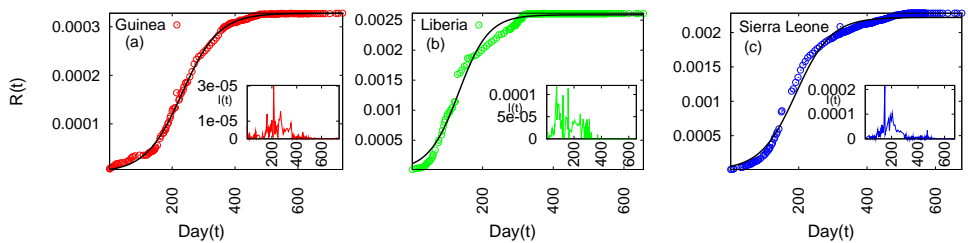}
\caption{Cumulative fraction of population infected and the fitted curve as a function of time (day) for  countries (a) Guinea, (b) Liberia and (c) Sierra Leone.
 Insets are for the fraction of newly infected population  as a function of time of each country.}
\label{fig:gls_fit}
\end{figure}

The  removed  population in the model in \cite{Khaleque:2013} was fitted to the form:
\begin{equation}
R(t) = \frac{a\exp(t/T)}{1+c\exp(t/T)},
\label{fit}
\end{equation}
where $a$, $c$ and $T$ depend on the values of $\delta$ and $q$. 
Note that the removed population in the model essentially corresponds to  the cumulative infected cases  since in the
model the infected agents were assumed to be removed immediately after being infected.   
This fitting form is used for the cumulative data of infected cases  and shows very good agreement 
 for Guinea (Figure \ref{fig:gls_fit}(a)),  while there is  fairly good agreement with the data of the other two countries (Figures. \ref{fig:gls_fit}(b),\ref{fig:gls_fit}(c)).  
Rescaling $R(t)$  such that it varies from $0$ to $1$, one can find out the goodness of fit. 
We performed the Kolmogorov-Smirnov test to evaluate the goodness of the fit for all the three 
sets of data. The values  are:    1.1294 
for Guinea (sample size $N_s = 217$),    
1.1070 for  Sierra Leone    ($N_s = 235$)
and  1.4959 for Liberia ($N_s = 233$).
Thus  the fittings are acceptable at the  level of significance $\alpha  = 0.10$ for Guinea and Sierra Leone and 
at $\alpha = 0.01$ for Liberia. 

From eq. (\ref{fit}),  one can show that a peak value for $I(t)$ will occur at $t_p = T \log(1/c)$.
The associated values of the exponents  $a,c$ and $T$ are found out for the three countries and the 
values of $t_p$ also extracted.  
We have plotted the data for $I(t)$ against $t$ in the  insets of the Figures \ref{fig:gls_fit}(a),~\ref{fig:gls_fit}(b) and \ref{fig:gls_fit}(c). 
We observe a lot of fluctuations and not a very clear peak in the data just as in the case of SARS \cite{Watts:2005}. Even then, the theoretically estimated values of $t_p$
 tally with a large value  of new cases occurring close to this time.  
 The exponent  values and $t_p$  are tabulated  in Table \ref{tab:expo}. 
 The  errors in the estimation of exponents are  ${\mathcal{O}}(10^{-6}$) for $a$, ${\mathcal{O}}(10^{-3}$) for $c$ and ${\mathcal{O}}(10^{0}$) for $T$ (beween 0.9 to 7.6 percent).
 One can see that $t_p$ is also directly proportional
 to the total duration, being least for Liberia and maximum for Guinea.

\begin{table}[ht]
\centering
\begin{tabular}{|l|l|l|l|l|l|}
\hline
Country &   $a$        &$  c$ & $  T  $& $  t_p  $&$a/c$\\ 
\hline
 Guinea     & 0.0000059$\pm$0.0000002  &  0.0182146$\pm$0.0006 &   57.9150$\pm$ 0.503& 231.98 & 0.0003239 \\ 
 \hline
 Liberia   &  0.0001125$\pm$0.0000085 & 0.0434763$\pm$0.0032 & 42.1957$\pm$ 1.043& 132.30& 0.0025876\\ 
 \hline
 Sierra Leone  & 0.0000549$\pm$0.0000041 & 0.0247653$\pm$0.0018 &  51.758$\pm$ 1.040 & 191.41&0.0022168\\  
\hline
\end{tabular}
\caption{\label{tab:expo} Exponents $a,c$ and $T$ for three different countries using the total population as normalization factor. }
\end{table}

\subsection{Results from the model}
The cumulative data for infected people has a sigmoid form  in general and has  been shown to have a form   given by eq (\ref{fit}) in a recent study as well \cite{Burg:2015}. To
establish that indeed the Euclidean network is an appropriate model responsible for the epidemic spreading, one should 
be able to reproduce from the model the consistent results and trends using suitable values of the parameters, at least qualitatively.  

Epidemic spreading on the  Euclidean model with the two parameters $\delta $ and $q$, already mentioned in the context of data analysis,   was first considered in \cite{Khaleque:2013}.
The model and simulation methods are given in detail in the Methods section.

\begin{figure}[ht]
\centering
\includegraphics[width=8.5cm]{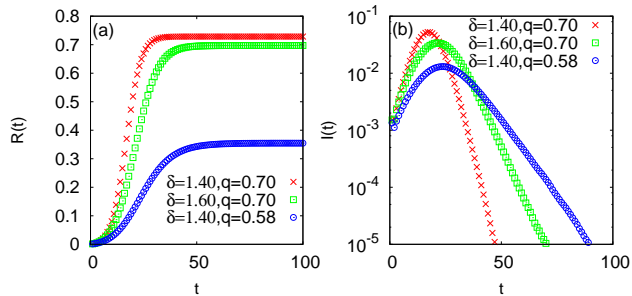}
\caption{(a) Fraction of population infected as a function of time (Monte Carlo time step) for different pairs of infection rate $q$ and $\delta$.
(b) Fraction of newly infected population as a function of time (Monte Carlo time step) for same pairs of infection rate $q$ and $\delta$.}
\label{fig:GSL_sim}
\end{figure}

\begin{table}[ht]
\centering
\begin{tabular}{|l|l|l|l|l|l|}
\hline
Parameters &   $a$        &$  c$ & $  T  $& $  t_p  $&$a/c$\\ \hline

  $\delta=1.4$, $q=0.58$ (Guinea)   & 0.011842 & 0.032342 &  7.09658& 24.35107&0.3661492\\ \hline

$\delta=1.4$, $q=0.70$ (Liberia)    & 0.005116  &  0.006974 & 3.39171& 16.8415& 0.7335818 \\  \hline

 $\delta=1.6$, $q=0.70$ (Sierra Leone)   &  0.011130 & 0.015718 & 5.10908& 21.2175&0.7081053 \\

\hline
\end{tabular}
\caption{\label{tab:expo_model} Exponents $a,c$ and $T$ for  different values of parameters. The countries to which the data correspond are  shown within the parenthesis.}
\end{table}

The behaviour of the network depends on the value of $\delta$. The network behaves as a small
world network for $\delta<1$ and as a regular one dimensional
lattice for $\delta>2$. For $1<\delta < 2$, it shows short range behaviour.
These properties of the network had been earlier detected by considering its network properties as well as 
critical phenomena on the network (see \cite{Arnab:2006,Khaleque:2013} and the references in these papers). 

The Ebola virus spreads through actual body
 contact and in most cases the infection occurred within family members. Hence the  underlying network must be short 
  ranged. 
Therefore to  get results comparable to the real data, 
one should use  a value of
$\delta$ larger than  $1$. 
Also, $\delta < 2$ is chosen as a real network is more connected  than a regular one.  
 The values of $q$ should be same in principle as it depends on biological  factors.   However, the value of  $q$ may be 
effectively altered using 
 control schemes like contact tracing,  quarantining the patient and  efficiently treating the disease. Such possibilities have not been directly included in the model. 
We will address this issue in the next section again.
 
We first discuss the case of Liberia and Sierra Leone. We note that the
saturation values are quite close while the saturation in Liberia has been 
reached earlier (Figure \ref{fig:gls}). We find that the same value of $q$ but a different
value of $\delta$ can indeed reproduce these features; the red and green curves in Figure \ref{fig:GSL_sim}(a) show the results
 for $\delta = 1.4$ (for Liberia) and $1.6$ (for Sierra Leone) while the $q$ values are same ($q=0.70$).


We next discuss the case for Guinea. It  has the lowest saturation value of the cumulative data for infected
 population while the disease is of duration slightly longer than that of Sierra Leone. 
This makes it quite apparent that one has to use a smaller value of $q$ to get data consistent with that of Guinea. 
We find that indeed one can get such values of $q$ keeping $\delta = 1.4$ such that the saturation 
value is smaller
while the duration is larger comparatively. 
We show the data by the blue curve in Figure \ref{fig:GSL_sim}(a) using $q=0.58$.

Of course these are some typical values which yield results comparable to the real data. A  range of values exist which 
more or less show the same behaviour. However, that range is not too large which would mean that the values are 
irrelevant. For $\delta$, this range is $\pm 0.05$ while for $q$ it is $\pm 0.02$. 

Figure \ref{fig:GSL_sim}(b) shows the data for $I(t)$ against $t$ from the Euclidean model. Again we find consistency, the red curve has a peak 
occurring earliest while the disease lasts for the shortest duration which corresponds to Liberia. The green curve shows a peak occurring at a later time and 
the duration is also longer. This we claim to correspond to Sierra Leone. The peak value is slightly less in height for the green curve compared to the red which
is also consistent with the real data (up to a multiplicative factor) if one takes the single spike occurring in  Figure \ref{fig:gls_fit}(c) inset to be spurious. The data for Figure \ref{fig:gls_fit}(a) inset is easily
comparable to the  red curve in Figure \ref{fig:GSL_sim}(b). The blue curve in Figure \ref{fig:GSL_sim}(b) corresponds to Guinea as the peak
 value occurs at a slightly larger time compared to the green curve while the duration is longest.
The quantitative values of $t_p$ shown in Table \ref{tab:expo_model} are also consistent with the real data. The errors in the estimates lie between 0.04 to 0.49 percent.  
The argument behind the choices of $\delta$ and $q$ are  discussed in the next section.

As in the real data, one can quote here the goodness of fit for $R(t)$ from the 
simulations. The  
Kolmogorov-Smirnov test for the three cases yield the largest 
errors  as  
0.1653 for 
$\delta =1.40$, $q=0.70$; 
0.3558 for 
$\delta =1.60$, $q=0.70$;
and 1.2261 for $\delta =1.40$, $q=0.58$. The sample size is 2000 for each 
and therefore the first two results are acceptable at level of significance 
$\alpha = 0.2$ and the last one at $\alpha = 0.05$.


\subsection{Comaparison of data by rescaling}

While qualitative
features of the data obtained from the model are quite similar to the
real data, the actual values of the fitting parameters $a$,   $c$ and $T$ (and consequently $t_p$) are quite different
(see Tables \ref{tab:expo} and \ref{tab:expo_model}).
It may be noted that $a/c$  corresponds to the saturation value of $R(t)$ and $c$ and $T$ determine the value of $t_p$.
The mismatch of the $t_p$ values is not surprising,  unit of time in the model is  just one Monte Carlo (MC) time step that has got nothing to do
with real time. On the other hand, the saturation values depend heavily on the normalization factor. The actual population
who are susceptible may form only a subset of the total population so saturation values can be different changing the values of $a$ and $c$. Nevertheless,
we find that the ratio of  of $a/c$  from the data and from the model for Sierra Leone and Liberia  are very close
which indicates that the fractions of susceptible population in these two countries were comparable while for Guinea
it was smaller. Indeed, Table \ref{tab:statistics} shows that the density of infected population were same for Sierra Leone and Liberia and order of magnitude smaller in Guinea.

\begin{figure}[ht]
\centering
\includegraphics[width=6.5cm]{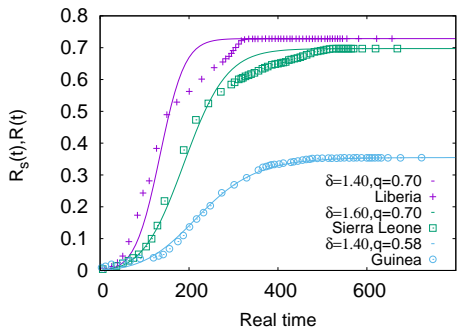}
\caption{ Solid curves are for the fraction of population infected (simulated data) $R(t)$ as a function of real time (rescaled Monte Carlo time step) for different pairs of infection rate $q$ and $\delta$. The rescaling factors for the MC time steps are 8.0 for Liberia, 9.0 for Sierra Leone and 9.2 for Guinea.
Data points  are for the rescaled fraction of population infected (real data) $R_s(t)$ as a function of real  time for the three countries Guinea, Liberia and
Sierra Leone.}
\label{fig:comp}
\end{figure}

\begin{table}[ht]
\centering
\begin{tabular}{|l|l|l|l|l|}
\hline
Country	&Total Cases &  Density of Infected Population & Lab-Confirmed Cases&	Total Deaths   \\
\hline 
Guinea & 	$3814$& 	$3.0 \times 10^{-4}$  &  $3358$& 	$2544$\\
\hline 
Sierra Leone &	$14124$ & $2.2 \times 10^{-3}$ & $8706$&	$3956$\\
\hline 
Liberia&	$10678$& 	$2.2 \times 10^{-3}$& $3163$&	$4810$\\
\hline 
Total	& $28616$&& 	$15227$& 	$11310$\\
\hline
\end{tabular}
\caption{\label{tab:statistics} Statistics of Ebola data for three different countries.}
\end{table}

However, one can still explore the possibility of rescaling the real data to obtain better quantitative agreement between the parameters. This 
may be possible by suitably choosing a normalization factor for each of the three data sets. Assuming the total population which has been removed at 
time $t$ to be $R_{tot}(t)$ and $\rho N_p$ the actual  population  susceptible to the disease, we calculate the density 
\begin{equation}
R_{s} (t)  = R_{tot}(t)/\rho N = R(t)/\rho,
\end{equation} 
where $R(t)$ is the density calculated earlier by dividing $R_{tot}(t)$ by the total population $N_p$  and $\rho$ a proportionality constant.
Taking  $R_{s} (t)$ to be the saturation value of $R(t)$ obtained from the model and equating it to  the saturation value of  $R(t)$  obtained 
from the real data, one can estimate the value of $\rho$. For example, for Guinea, the saturation value of $R(t)$ is $0.000328$ from the data 
and from the model it is 0.354720. Hence $\rho  \approx 9.24 \times 10^{-4}$. Similarly for Sierra Leone and Liberia it is  $3.27 \times 10^{-3}$ 
and $  3.58 \times 10^{-3}$ respectively.  On the other hand, one can compare the $t_p$ values from
real data and the model and we find that  the timescales in the real data are approximately $8-9.2$ times the timescales in the Monte Carlo simulations. 
Hence we also rescale the time for the results obtained for the model. The rescaled data $R_s(t)$ and $R(t)$ are plotted against ``real time'' in Fig. \ref{fig:comp} and show an excellent agreement
for Guinea and a reasonably good agreement for Sierra Leone.   
The agreement for Liberia is not that good, however, the data for Liberia are somewhat irregular and it is difficult to fit them 
with a smooth function very accurately as already noted. 
Particularly for Liberia and Sierra Leone we find that before saturation, there is a slower increase in $R(t)$; this might be due to an enhancement
in the treatment and preventive measures against the disease. 

One can similarly rescale the newly infected density $I(t)$, however, the data being too noisy, we do not attempt that. Nevertheless, we find that the peak values of the newly infected density $I(t)$,  when scaled by $\rho$    shows order of magnitude agreement with the model data. 

Further, we have fitted $R_s$ using equation \ref{fit} and present the value of the parameters in Table \ref{tab:expo2}. The values from the model and the real data are easily comparable now showing order of magnitude agreement for most of them.

\begin{center}

\begin{table}[ht]

\centering

\begin{tabular}{|l|l|l|l|l|l|}

\hline

Country &   $a$        &$  c$ & $  T  $& $  t_p  $&$a/c$\\ 

\hline

 Guinea     & 0.0077722 $\pm$ 0.00070 &  0.0214467 $\pm$ 0.00189 &  59.5716 $\pm$ 1.33& 228.88 & 0.362399\\ 

 \hline

 Liberia   &  0.0320618  $\pm$ 0.00464 & 0.0443399 $\pm$ 0.00636 & 42.5638  $\pm$ 2.02& 132.62& 0.723091\\ 

 \hline

 Sierra Leone  & 0.0274868  $\pm$ 0.00450 & 0.0387566  $\pm$ 0.00603 & 57.3669  $\pm$ 2.41 &186.47&0.709215\\  

\hline

\end{tabular}

\caption{\label{tab:expo2} Exponents $a,c$ and $T$ for rescaled data $R_s(t)$ for three different countries. The values of $a$ and $c$ can be compared to those appearing in Table 2.  $t_p$ values 
are approximately 8-9 times compared to the $t_p$ values obtained in the Monte Carlo simulations (Table 2).}

\end{table}

\end{center}

\section{Discussion}

In this section we justify the choice of the parameters used in the model  to 
obtain the results consistent with the real data. 
One can of course attempt to get a  full calibration of $a$ and $c$ for given values of $\delta$ and $q$ 
so that the choice of $\delta$ and $q$ are automatically obtained from this calibration, however, we have refrained from doing so as it 
involves a  huge computational calculation.

We have already justified the choice of $\delta$ between $1$ and $2$ in the last section. We have used a larger value 
of $\delta$ for Sierra Leone and a smaller value of $q$ for Guinea to get the consistency.  
To justify why $\delta$ should be larger for Sierra Leone we note 
the following. 
Sierra Leone and Liberia are comparable in size but the   density of population is   much higher in the former. 
The density of population is $79.4$/km$^2$ and $40.43$/km$^2$ respectively for these two countries \cite{Popu}. 
Hence the number of neighbours within the same distance is larger for Sierra Leone which implies a larger value of $\delta$ effectively (more short ranged).  

On the other hand, the population densities of Guinea ($40.90$/km$^2$) \cite{Popu1} and Liberia are quite close so
that one should use the same $\delta$ value. 
However, we need  to justify why a smaller value of $q$ is able to reproduce
the data for Guinea. A smaller value of $q$ indicates less infection probability which 
is possible if proper medical care and control measurements are taken. 
This is indeed true as we find from several documents that 
the disease was tackled most effectively in Guinea. Table  \ref{tab:statistics} clearly shows that 
the maximum percentage of  cases for Guinea were laboratory-tested which indicates that the process
of contact tracing and treatment were more efficient. 
This is supported by the fact that in  Guinea, about $56$ contacts per infected person were traced compared to $23$ in case
of Sierra Leone 
\cite{G_sl}. 
We find from 
\cite{Guinea} that MSF treated the largest number  
of reported cases in Guinea, in Sierra Leone the  minimum out of reported cases. Thus most cases in Sierra Leone, 
even when reported, had  received less attention while in Liberia, a large number is not confirmed or reported at all. 
Apparently,  medical centers by international organisations have also been set up much earlier in  Guinea as 
it was the epicenter of the disease and the disease started as early as in 2013 December. However, 
later activities could control the disease in Liberia and Sierra Leone as well, and the final number 
of deaths had been far less than initially anticipated. 
 We also note a curious fact - though Guinea may have recorded the minimum number of cases, 
yet the disease spanned a longer duration compared to Liberia. Further analysis, beyond the scope of the present paper, may be able to 
explain this.  

Although we have shown that by rescaling the real data by $\rho N_p$ and the MC time by a suitable factor, one can get fairly good agreement between the real data and the simulated data,
it has to be emphasized that the rescaling is somewhat manipulated by the results of the model. The ratio of the saturation values 
for the real data and the simulated data corresponds to the factor $\rho$. In principle one should incorporate more factors 
in the model to fit the real data independently. 
However,  at the 
present stage qualitative consistency is what we emphasize on. 
To achieve quantitative consistency one needs to introduce more parameters making the model complex.
These parameters may be related to features like inhomogeneity, mobility, 
more general initial conditions to name a few.  We have  made simple assumptions like homogeneity, i.e., uniform number of contacts for all agents. 
The initial condition has been taken to be identical: the disease
commences with only one infected person. 
Our assumption that  agents are  immobile is supported by \cite{Burg:2015} in which it is  argued that
migration did not play a role in the spreading. 
Even so, this simple model is able to yield data which is consistent with real data and we conclude that it captures the basic mechanism of 
the epidemic spread. 
The effect  of the Ebola outbreak has been devastating in the West African countries. Apart from
the human losses, economic loss has also been considerable \cite{Fighting_ebola}. 
The present study shows that the Euclidean model  can be treated as a basic starting point  and  can be further developed by 
adding other features.  This will 
make it very useful and   important  for  making accurate predictions.  
\section{Methods}
\label{methods}

\subsection{How the database was handled}
We consulted the Ebola data for the number of cases detected in  the three countries Guinea, Liberia and
Sierra Leone in West Africa (The Centers for Disease Control and Prevention (CDC) \cite{CDC}).
 The data is available from 25th March 2014 to 13th April 2016  
 at the time interval of a few days. The data is noisy and contains obvious errors as sometimes the cumulative data is shown to decrease which is unphysical.
The first available data is from March 2014 when Guinea was already struck with the disease for some time (first case in Guinea reported in December 2013) 
such that the data for the initial period is missing. 
For Liberia and Sierra Leone, the data for initial stage are available, however these are  sparse and unreliable;
 often the data for number of death 
 exceeds the number of cases.   For this reason, 
the data has been analysed from the date when the number of 
cases detected is at least $50$ for each country. 
Even then the errors cannot be fully avoided as  for  very late stages, the data  being rare, also become somewhat unreliable.
Hence, the entire data set has to be handled carefully.

In Table \ref{tab:statistics}, a summary of the statistics of the Ebola data is presented and
one can immediately note that all cases could not have been confirmed in the laboratory in the case of Liberia
where number of deaths exceeds the laboratory confirmed cases. Obviously many cases were unreported.
For Guinea, these two figures are closest and the data for Guinea is in fact the cleanest one.


Another point needs to be mentioned. The disease has been officially declared over on 1st June 2016 for Guinea, 9th June 2016 for Liberia and 
 17th March 2016 for Sierra Leone \cite{Declare}. But one can see from Figure \ref{fig:gls}
 that the cumulative data shows a saturation over fairly long period of time.
Apparently a few stray cases delayed the declaration of the disease being over. For Liberia, for example, the disease was originally declared to be
over as early as in May 2015 but two small flare-ups were reported later. However the cumulative data is hardly affected by the later cases.

\subsection{Model and Simulation method}
In the Euclidean model, the nodes of the network are assumed to occupy the sites of a chain of length $N$. 
We generated  random  long range bonds by connecting nodes located at a distance $l$ along the chain with a probability 
$P(l) \propto l^{-\delta}$; the probability is normalised by making the total probability equal to unity.
Once $N/2$ such bonds are constructed, the network is completed. The average degree of each node is three and it 
is expected that the inhomogeneity of the degree distribution is negligible. 
The disease spreading process is then simulated by assuming a single infected agent at any randomly chosen site in the beginning.
All the neighbours are likely to be infected with a probability $q$ in the next step. One generates a random number between $0$ and $1$, 
if it is less than $q$, the agent is taken to be infected. From the agents who are infected in the second step, 
the disease spreads to their neighbours and the process continues. 
Infected people   
are  removed within one unit of time, with the assumption that they are either dead or cured,  and they can infect the susceptible agents during this one time step only.
The dynamical evolution stops
when there is either no infected agent at a particular step or when all of them have died. Several configurations are considered and the 
dynamical variables averaged.   

In the present simulation, for the same network, the initial choice of infected site was repeated 
400 times and the quantities averaged. A secondary averaging is made by considering 100 different 
network configurations. 
The number of nodes $N$ and the total number of edges were kept fixed  for
any value of $\delta$ and $q$ in the different realisations.
Periodic boundary condition has been used in the simulation. Systems with size $N = 2^{11}$ has been considered. 
\vskip 0.5cm
Acknowledgement: PS acknowledges support from CSIR grant.

\end{document}